\documentclass{article}
\usepackage{spconf,amsmath,graphicx}

\usepackage{enumitem}
\setlist{nosep, leftmargin=14pt}

\usepackage{mwe} 

\usepackage{graphicx}
\usepackage{amssymb}
\usepackage{amsmath}
\usepackage{amsfonts}
\usepackage{bbm}
\usepackage{psfrag}
\usepackage{xcolor}
\usepackage{fullpage}
\usepackage{epstopdf}
\usepackage{graphicx}
\usepackage{times}
\usepackage{commath}
\usepackage{thmtools}
\usepackage{thm-restate}
\usepackage{times}
\usepackage{graphicx}
\usepackage{dirtytalk}
\definecolor{ddarkbrown}{rgb}{0.5,0.2,0.05} \definecolor{bbluegray}{rgb}{0.05,0,0.5}

\usepackage[colorlinks,citecolor=bbluegray,linkcolor=ddarkbrown,urlcolor=blue,breaklinks]{hyperref}

\usepackage{subfig,float} 

\usepackage{mathtools}

\usepackage{algorithm,algcompatible,algpseudocode}
\usepackage{multicol,lipsum}
\algnewcommand{\Inputs}[1]{%
	\State \textbf{Inputs: \:}{#1}
}

\algnewcommand{\Output}[1]{%
	\State \textbf{Output: \:}{#1}
}
\algnewcommand{\Initialize}[1]{%
	\State \textbf{Initialize: \:}{#1}
}

\algnewcommand{\IIf}[1]{\State\algorithmicif\ #1\ \algorithmicthen}
\algnewcommand{\EndIIf}{\unskip\ \algorithmicend\ \algorithmicif}

\let \oldsection \section
\renewcommand{\section}{\vspace{3ex plus 1ex}\oldsection}

\newcommand{\BEAS}{\begin{eqnarray*}}
	\newcommand{\EEAS}{\end{eqnarray*}}
\newcommand{\BEA}{\begin{eqnarray}}
\newcommand{\EEA}{\end{eqnarray}}
\newcommand{\mb}{\mathbb}
\newcommand{\BEQ}{\begin{equation}}
\newcommand{\EEQ}{\end{equation}}
\newcommand{\BIT}{\begin{itemize}}
	\newcommand{\EIT}{\end{itemize}}
\newcommand{\BNUM}{\begin{enumerate}}
	\newcommand{\ENUM}{\end{enumerate}}

	\newcommand{\D}{\mathcal{D}}
		\newcommand{\U}{\mathcal{U}}
	\newcommand{\Pm}{\mathcal{P}}
	
	\newcommand{\mr}{\mathrm}

\newcommand{\BA}{\begin{array}}
	\newcommand{\EA}{\end{array}}

 \numberwithin{dummy}{section}

\numberwithin{mythm}{section}
\numberwithin{mydef}{section}
\numberwithin{myprop}{section}
\numberwithin{mylem}{section}
\numberwithin{mycor}{section}

\usepackage{enumitem}
\setlist{nosep, leftmargin=14pt}

\usepackage{mwe} 

\title{Fast Gradient Methods for Data-Consistent Local Super-Resolution of Medical Images}
\name{Junqi Tang, Guixian Xu, Jinglai Li\thanks{Emails: Junqi Tang (j.tang.2@bham.ac.uk), Guixian Xu (gxx422@student.bham.ac.uk), Jinglai Li (j.li.10@bham.ac.uk)}}
\address{School of Mathematics, University of Birmingham}
\begin{document}
%
\maketitle
\begin{abstract}
  In this work, we propose a new paradigm of iterative model-based reconstruction algorithms for providing real-time solution for zooming-in and refining a region of interest in medical and clinical tomographic images. This algorithmic framework is tailored for a clinical need in medical imaging practice that after a reconstruction of the full tomographic image, the clinician may believe that some critical parts of the image are not clear enough, and may wish to see clearer these regions of interest. A naive approach (which is highly not recommended) would be to perform the global reconstruction of a higher resolution image, which has two major limitations: first, it is computationally inefficient, and second, the image regularization is still applied globally, which may over-smooth some local regions. Furthermore, if one wishes to fine-tune the regularization parameter for local parts, it would be computationally infeasible in practice for the case of using global reconstruction. Our new iterative approaches for such tasks are based on jointly utilizing the measurement information, efficient up-sampling/down-sampling across image spaces, and locally adjusted image prior for efficient and high-quality post-processing. The numerical results in low-dose X-ray CT image local zoom-in demonstrate the effectiveness of our approach.
\end{abstract}
\section{Introduction}

Medical tomographic imaging is one of the pillar areas in healthcare. Unlike other application scenarios in general computational imaging, medical imaging professionals pay significant more attention and care in critical areas, aka, regions-of-interest (ROI) \cite{bubba2016roi}. Quite often, observing the abnormal (for example, a tumor or a tiny crack in some bone) in a patient is much more critical than reconstructing a clean overall image \cite{antun2020instabilities}. Traditional and current state-of-the-art model-based iterative reconstruction algorithms, even plug-and-play / regularization-by-denoising schemes \cite{venkatakrishnan2013plug,romano2017little} with kernel-based \cite{dabov2007image, tachella2020neural} or deep learning-based image priors \cite{zhang2017beyond,zhou2024deep,tang2025practical,tan2024provably}, are all tailored for global regularization and reconstruction of images and do not have a mechanism to focus on improving the reconstruction quality of local parts.

In this work, we propose a practical solution for super-resolution and refine locally critical parts of tomographic images reconstructed from measurements given by medical imaging systems such as CT, MRI and PET, etc. Such imaging systems can generally be expressed as
\begin{equation}\label{model}
   b \approx A x^\dagger,
\end{equation}
where $x^\dagger \in \mb{R}^d$ denotes the ground truth image (vectorized), and $A \in \mb{R}^{n \times d}$ denotes the forward measurement operator, while $b \in \mb{R}^n$ denotes the noisy measurement data. A classical way to obtain a reasonably good estimate of $x^\dagger$ is to solve a composite optimization problem:
\begin{equation}\label{1st-stage}
    x_1^\star \in \arg\min_{x \in \mb{R}^d} f(b, Ax) + \lambda g(x),
\end{equation}
where data fidelity term $f(b, Ax)$ is typically a convex function (one example would be the least squares $\|b - Ax\|_2^2$), while $g(x)$ being a regularization term, for example the total-variation (TV) semi-norm, and $\lambda$ denotes the regularization parameter controlling the strength of the regularization effect \cite{chambolle2016introduction}. 

Regularization, both classical variational regularization and more advanced learning-based regularization, is applied to the high-dimensional image space globally for overall good reconstruction quality. In the process of iterative reconstruction, the imaging quality of critical small regions of the image can often be compromised. These critical areas, such as infested parts and tumors, are usually dependent on the patient and therefore cannot be known a priori. It would be desirable if we could perform a fast refinement and super-resolution postprocessing step on critical parts pinpointed by clinicians in the reconstructed image by solving (\ref{1st-stage}), utilizing a locally adjusted regularization. In this work, we propose a class of iterative reconstruction algorithms for these types of applications.

Alternatively, one may consider reconstructing an extra-high-resolution image instead. However, this would introduce a significant amount of extra computation, and meanwhile it still does not address the limitation of global regularization and would make the inverse problem more ill-posed. In clinical practice, it is highly desirable to provide medical doctors a path of images with a grid of different regularization parameters and let them decide which of the images is the most likely to be closest to the truth. In this situation, a global reconstruction of a sequence of images would be very computationally costly compared to our local reconstruction approach.

\section{Data-Consistent Local Zoom-in}

In this section, we present our algorithmic framework for achieving efficient data-consistent local zoom-in for medical images.

\subsection{Optimization Framework}

Ideally, a successful algorithm for such a local zoom-in and refinement should take into account the measurement data, whole reconstructed image from the first stage, and locally adjusted image prior. Let $x_z$ denote the part of $x_1^\star$ that clinicians want to zoom in and refine, and $x_o$ be the complement of $x_z$, $A_z$ be the block of $A$ associated with the coordinates of $x_z$, and $A_o$ being the complement of $A_z$, while $v \in \mb{R}^m$ be the variable for the desired high resolution image block. Denote 
\begin{equation}
    b_z = b - A_o x_o,
\end{equation}
and $ \D(\cdot)$ being some down-sampling operator, and suppose that we use least-squares data fit, we write our program as:
\begin{equation}\label{obj}
    x_s^\star = \arg\min_{v \in \mb{R}^{m}} \|b_z - A_z \D(v)\|_2^2 +  \lambda g(v),
\end{equation}
Then a proximal gradient-like method for approximately solving (\ref{obj}) can be written as:
\begin{equation}
    x_{k+1} =  \Pm[x_k - \eta \cdot \U(A_z^TA_z \D(x_k) - A_z^Tb_z)],
\end{equation}
where \[\Pm(\cdot):=\mr{prox}_{\lambda g}^\eta(\cdot) = \arg\min_{x} \frac{1}{2\eta}\|x - \cdot\|_2^2 + \lambda g(x),\] $\eta$ being the step size, $\U(\cdot)$ being some up-sampling operator. We present our generic algorithmic framework for data-consistent local zoom-in as:
 \begin{eqnarray*}
 && \mathrm{\textbf{Local Zoom-in with Fast Gradients (LZFG)}}\\&& - \mathrm{Initialize}\ x_0\in \mb{R}^m \ y_0 \in \mb{R}^m \ a_0 = 1\\
 &&\mathrm{For} \ \ \ k = 0, 1, 2,...,  K\\
&&\left\lfloor
\begin{array}{l}
x_{k+1} \leftarrow  \Pm[y_k - \eta \cdot \U(A_z^TA_z \D(y_k) - A_z^Tb_z)]\\
a_{k+1} \leftarrow (1 + \sqrt{1 + 4a_k^2})/2;\\
y_{k + 1} \leftarrow x_{k+1} + \frac{a_{k} - 1}{a_{k+1}} (x_{k+1} - x_{k})
\end{array}
\right.
 \end{eqnarray*}
 Our algorithm here can be seen as an extension of the FISTA algorithm \cite{2009_Beck_Fast}, with additional up-and-down sampling elements. 

 \subsection{Exploiting local strong-convexity: Gradient-restart for accelerating LZFG}

 Observing that the local zoom-in optimization task is in fact more well-conditioned than the full image reconstruction, we believe that there is a chance for us to further accelerate LZFG via adaptive restart schemes for accelerated gradient methods. Here we extend the gradient-restart proposed in \cite{o2015adaptive} and obtain an improved version of LZFG:
  \begin{eqnarray*}
 && \mathrm{\textbf{LZFG with adaptive restart (R-LZFG)}}\\&& - \mathrm{Initialize}\ x_0\in \mb{R}^m \ y_0 \in \mb{R}^m \ a_0 = 1\\
 &&\mathrm{For} \ \ \ j = 0, 1, 2,...,  J\\
&&\left\lfloor
\begin{array}{l}
\mathbf{Execute\ LZFG\ until:}\\ \langle \U(A_z^TA_z \D(y_k) - A_z^Tb_z), x_{k+1} - x_k\rangle > 0 \\
x_0 \leftarrow x_{k+1},\ y_0 \leftarrow x_{k+1},\ \theta_0 \leftarrow 1
\end{array}
\right.
 \end{eqnarray*}
where we restart the momentum of LZFG whenever we detect $\langle \U(A_z^TA_z \D(y_k) - A_z^Tb_z), x_{k+1} - x_k\rangle > 0$, which indicates the approximate gradient direction is way too different to the momentum direction. Inheriting from gradient restart, R-LZFG is able to adaptively adjust the momentum to local strong-convexity of our zoom-in optimization problem, and hence provide faster and stable convergence, with negligible computational overhead (one evaluation of inner-product per iteration). As we will see in our experiments, such a restart scheme is indeed able to further improve the performance of LZFG.

\subsection{Computational advantages over repeating the full-scale global reconstruction}

In practical programming of the reconstruction algorithms for tomographic imaging such as CT and PET, we usually do not have direct access to the block operator $A_z$, but to the full forward operator $A$. This issue can be fixed easily and efficiently for CT and PET reconstruction -- in practice, we just need to construct a zero image of size $d$ and place the vector $\D(y_k)$ at the location of the region of interest (ROI), and then apply $A$ to this sparse image $v_s$ (since ROI is usually small), which is efficient. Since for CT/PET, the forward operator $A$ is a radon transform, $Av_s$ is also a sparse vector (also due to the fact that in each view only a small fraction of measurement can hit the ROI), hence the computation of $A^TAv_s$ is efficient -- $O(nm)$ complexity instead of $O(ndq)$ where $q$ is the upscaling factor. Meanwhile, the block backprojection term $A_z^T b_z$ in the gradient can also be easily pre-computed in the same way (compute $A^T b_z$ and then take the block corresponding to the ROI). After computing the gradient, we just need to take the partial gradient regarding the ROI and perform the update.

Our approach not only reduces the computational complexity of the gradient evaluation, but also significantly reduces the computational burden of the proximal operators. Instead of computing the proximal operators on the full high-resolution dimension, we only need to compute it in the local regime, which is tiny compared to the full dimension.

\section{Numerical Experiments}

 \begin{figure}[t]
   \centering
    {\includegraphics[width= .235\textwidth]{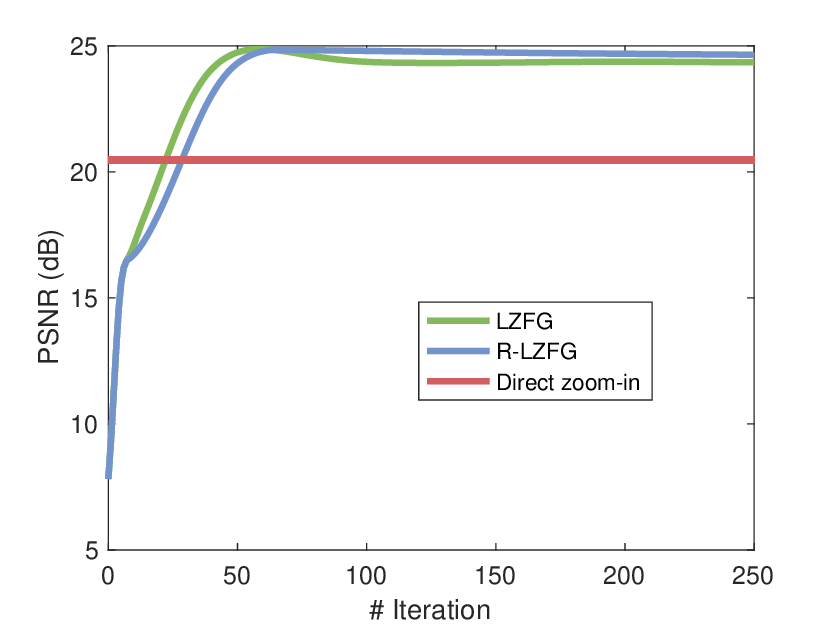}}
    {\includegraphics[width= .235\textwidth]{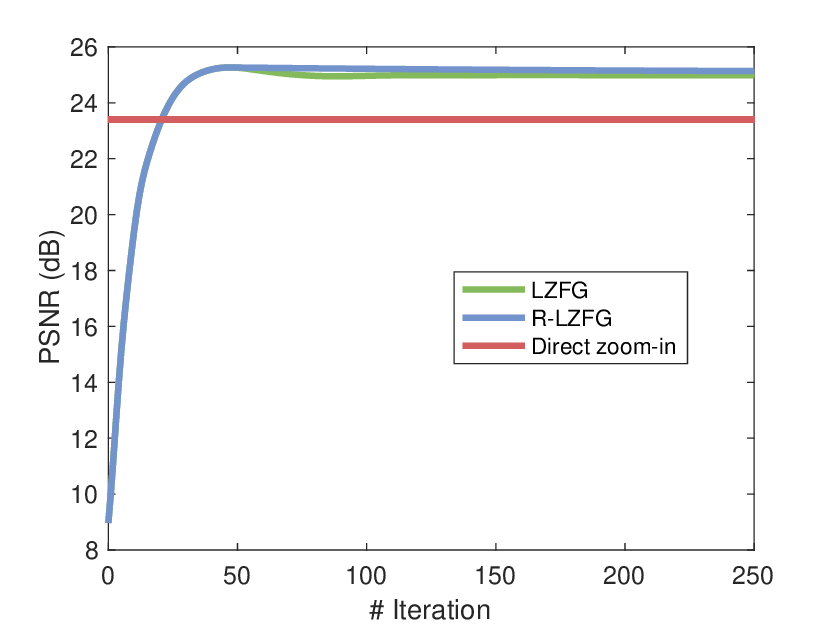}}
   \caption{Quantitative results for the 4 local zoom-in experiments on CT images. The first plot corresponds to the first example presented in Fig.\ref{e2}, and the second plot corresponds to the second example presented in Fig.\ref{e4}. From the PSNR results we can observe that our proposed data-consistent local zoom-in method LZFG achieves significantly improved reconstruction accuracy compare to the naive approach which directly zoom-in the first-stage reconstruction without utilizing the measurement data.}
   \label{f1}
\end{figure}
In this section, we present some preliminary results for the proof-of-concept. We consider an example of a low-dose fan-beam CT imaging system, and also an example of sparse-view fan-beam CT imaging system where we have reconstructed a head image and a chest image, with TV regularization. The measurements we simulate are corrupted with Poisson noise:
\begin{equation}
    b \sim \mathrm{Poisson}(I_0 e^{-Ax^\dagger}) 
\end{equation}
and we take the logarithm for the linearization of the measurements. We choose a low measurement energy with $I_0 = 2 \times 10^{3}$ and the forward operator $A \in \mathbb{R}^{92160 \times 65536}$ for low-dose CT example; while $I_0 = 2 \times 10^{4}$ and  $A \in \mathbb{R}^{13680 \times 65536}$ for sparse-view CT example. We run all the experiments here in MATLAB R2020b. Given the measurement and full operator, we solve a TV-regularized least-squares optimization problem by FISTA and get a global estimate of the ground truth. As we can observe, the reconstructed images have a reasonable global quality but could oversmooth some of the local areas which may need to be refined and zoomed in.  In this example, we seek to zoom-in 4 times larger a $50$ by $50$ block of a $256$ by $256$ image. Hence, the global iterative superresolution of the whole image will be at least 25 times more computationally expensive than our approach in this setting. Direct zoom-in on this block would often have poor performance since it does not re-utilize the measurement data. We then test our LZFG and R-LZFG, both using TV-regularization. We use the MATLAB \texttt{imresize} function as the up-sampling and down-sampling operators, with the default \texttt{bicubic} interpolation. 

\begin{figure}[t]
   \centering

    {\includegraphics[width= .49\textwidth]{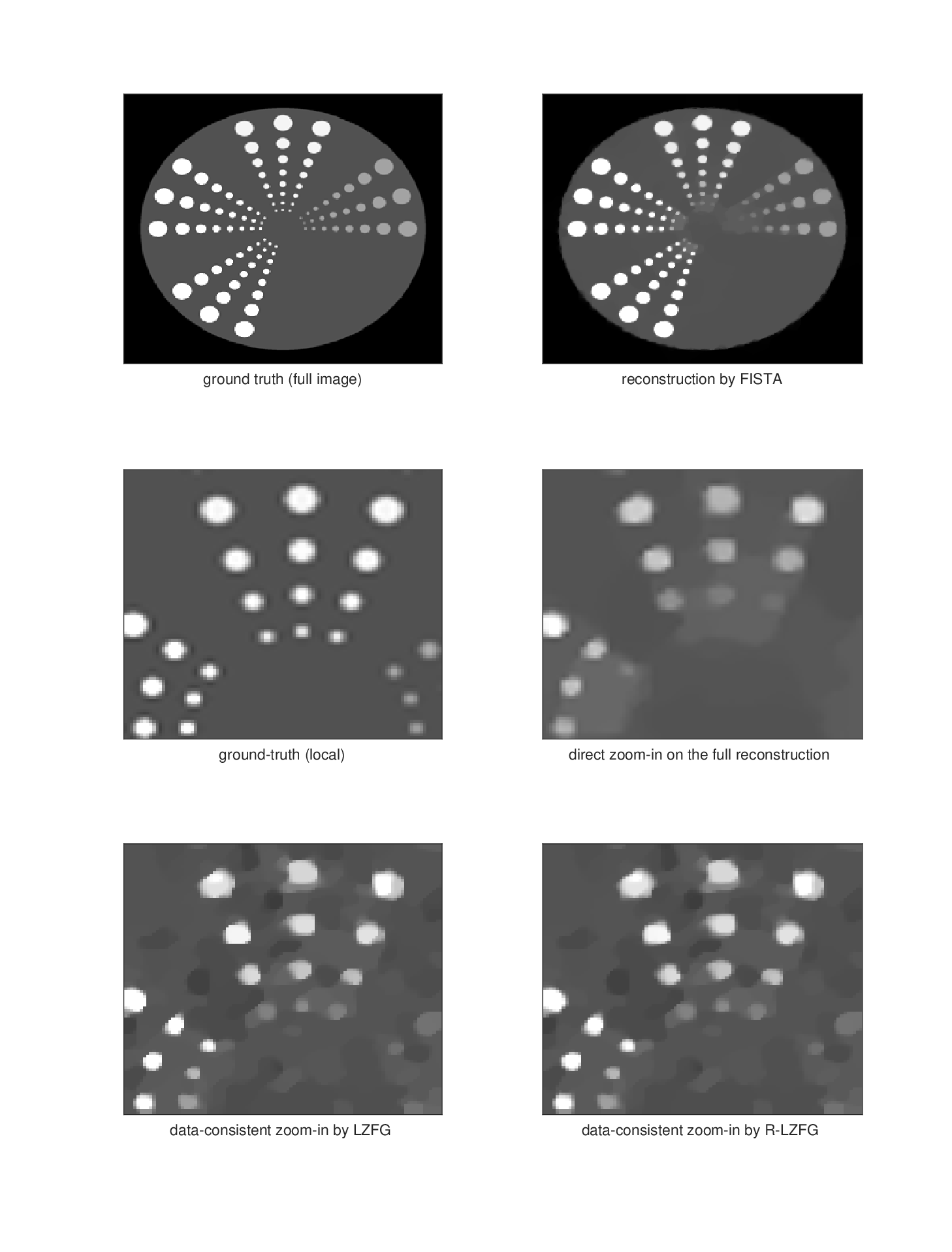}}

   \caption{Local zoom-in results for low-dose fan-beam CT image (example 1 with $I_0 = 2 \times 10^{3}$ and $A \in \mathbb{R}^{92160 \times 65536}$)}
    \label{e2}
\end{figure}

\begin{figure}[t]
   \centering

    {\includegraphics[width= .495\textwidth]{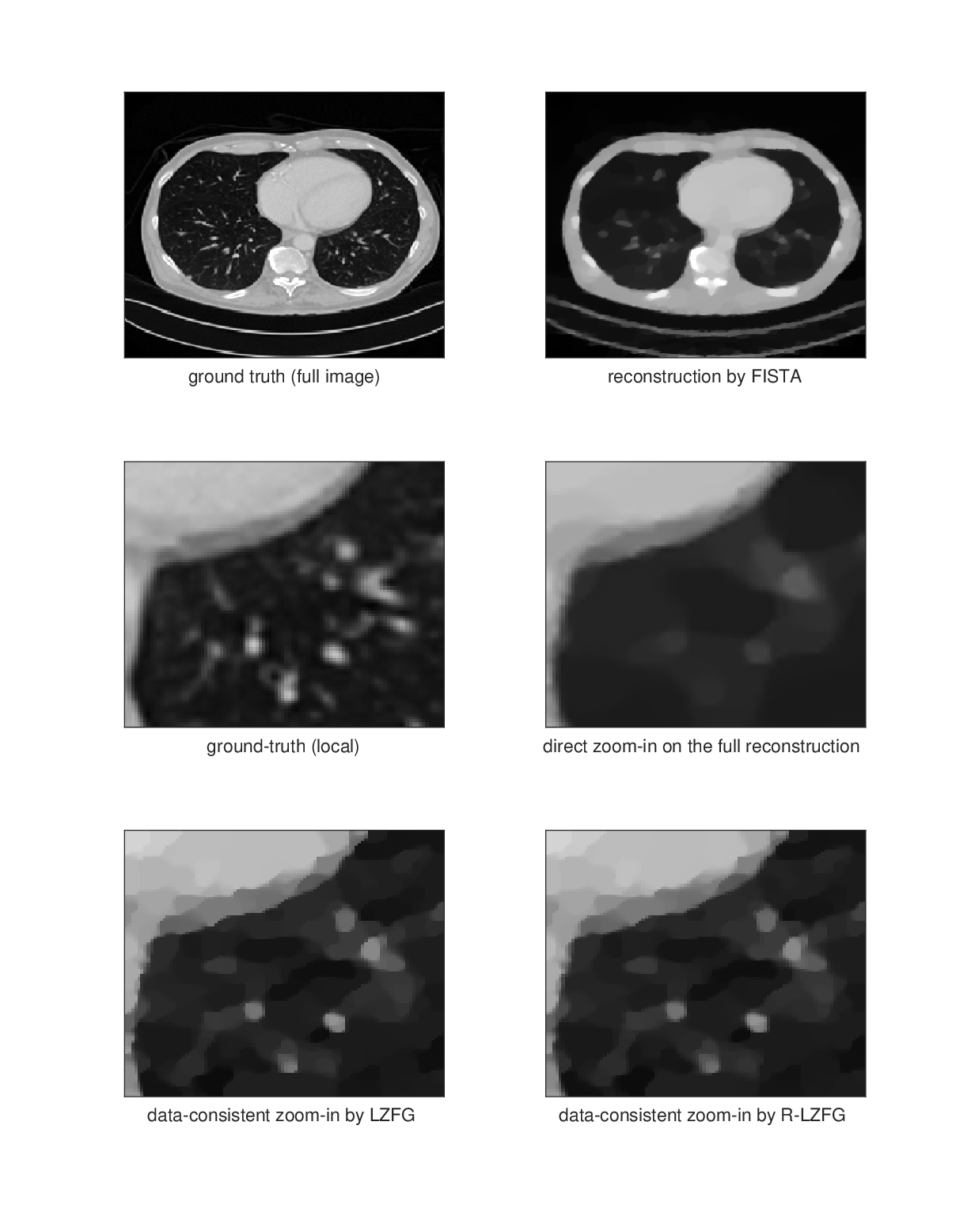}}

   \caption{Local zoom-in results for sparse-view fan-beam CT image (example 2 with $I_0 = 2 \times 10^{4}$ and $A \in \mathbb{R}^{13680 \times 65536}$)}
   \label{e4}
\end{figure}

In Figures \ref{e2} and \ref{e4} we present two examples of local zoom-in. We can observe from the numerical results that our method can indeed recover the details of the local blocks in high quality. The direct zoom-in from the first-stage reconstruction fails to recover local details missed from the first-stage reconstruction since it does not utilize the measurement data, demonstrating the importance of data consistency in local zoom-in of medical images. We present the convergence curves of LZFG and R-LZFG in Figure \ref{f1} reporting the PSNR results towards the zoomed-in ground-truth image. From the PSNR result, we can observe that our methods provide significantly improved reconstruction accuracy (around 2 to 5 dB better) compared to the direct approach, which does not take into account the measurement data. Meanwhile, we also observe that the R-LZFG with adaptive restart can further improve the convergence performance in the final stage of the LZFG as shown in Figure \ref{f1}, leading to a slight increase in PSNR within our computational budget.

\section{Conclusion}


In this work, we propose a new paradigm of iterative model-based reconstruction algorithms tailored for the practical need of clinicians, to provide a real-time solution for local superresolution and refining regions of interest of medical images, which may be unclear with missing details due to oversmoothing from the classical global reconstruction. Our LZFG methods are based on the proximal gradient methods, with up-sampling and downsampling operators allowing the algorithms to translate between high-resolution and low-resolution image spaces and approximate the gradient in the high-resolution image space using the low-resolution one. The numerical results demonstrate that our methods provide much better local reconstruction with improved imaging on details, compared to the naive zoom-in which does not re-utilize the measurement data.

In clinical practice, we would advise medical imaging practitioners to pre-define a range of regularization parameters (or parameters of the denoiser), and provide the clinicians a number of refined local images reconstructed under a grid of parameters, and let the clinicians decide which image to choose. Computing a regularization path with our scheme on local regions of interests (pin-pointed by clinicians after checking the first-stage reconstruction) will be much more efficient (orders of magnitude) than the brute-force approach which computes a regularization path of global reconstructions.

In this conference version, we focus purely on the optimization framework. Our approach can easily be extended to joint-force with deep learning-based priors via PnP/RED methods and stochastic gradient schemes \cite{ehrhardt2025guide}, and we leave the full study of these extensions in our journal version of this work. 


\bibliographystyle{IEEEbib}
\bibliography{strings,main}
\end{document}